\documentclass[10pt,twocolumn]{article}
\usepackage{geometry} 
\usepackage{color}
\usepackage{graphicx}
\pdfoutput=1
\title{Random drift versus selection in academic vocabulary: an evolutionary analysis of published keywords}
\author{R. Alexander Bentley\\Anthropology Department, Durham University\\Durham DH1 3HN UK\\r.a.bentley@durham.ac.uk}

\begin{document}
\maketitle

\begin{abstract}
The evolution of vocabulary in academic publishing is characterized via keyword frequencies recorded the ISI Web of Science citations database. In four distinct case-studies, evolutionary analysis of keyword frequency change through time is compared to a model of random copying used as the null hypothesis, such that selection may be identified against it. The case studies from the physical sciences indicate greater selection in keyword choice than in the social sciences. Similar evolutionary analyses can be applied to a wide range of phenomena; wherever the popularity of multiple items through time has been recorded, as with web searches, or sales of popular music and books, for example.
\end{abstract}

\section{Introduction}
Ideally, science is the systematic process of testing multiple hypotheses, but  as practiced by real people, it is also distinctly social. Within complex collaboration networks, academics compete for citations, particularly in our modern era of online citation databases that can `summarize' an academic's career at a single command  \cite{Guimera_etal_2005, Bentley_2006, Simkin_Roychowdhury_2007, Wuchty_etal_2007}. They are therefore prone to copy ideas, and particularly buzzwords, from one another \cite{Bentley_2006, Simkin_Roychowdhury_2003}. 
\\ \indent Diverse opinions exist as to what constitutes trendy ideas versus more meaningful research paradigms; the challenge is to evaluate this by some objective means. In other realms of fashion, ranked lists are increasingly a part of our world; from universities to Internet searches, downloads, book and music sales. Correspondingly, the design of algorithms needed to track `what's hot and what's not' has itself become a hot topic in computer science \cite{Hayes_2008}. Indeed, as journals are now ranked by their impact factor -- increasingly a subject of study \cite{Hirsch_2006, Stringer_etal_2008} -- there is no reason why we cannot look at academic buzzwords the same way: rank them in order of popularity from year to year, and track the comings and goings of `what's hot' on such lists.
\\ \indent As the science of how attributes are passed on and modified through time \cite{Nowak_2006}, 
evolutionary theory is an ideal means to model these aspects of scientific process \cite{Hull_2001}. Previous work using evolutionary models has shown, counter-intuitively, that many patterns of change in cultural choices over time can be explained as random drift; i.e.  the effect of chance on what happens to be copied, together with the occasional appearance of innovations \cite{Bentley_etal_2004, Shennan_Wilkinson_2001, Hahn_Bentley_2003}. Meaningful selection, as opposed to random copying, occurs when such choices are made on the basis of something inherent to the choice itself \cite{Shennan_2008}  - as with a `better mousetrap' for example, or something inherently preferable to human tastes. 
\\ \indent In knowledge production, ideas are not always adopted out of inherent superiority, but often merely because others are using those ideas. In either case, the transmission process is evolutionary; predominantly one of adopting what others have done, with creative modifications contributing new ideas that eventually replace old ones through being adopted. `Ideas' of course is a nebulous description, so this study focuses specifically on the evolution of keyword use in academic publishing.
\\ \indent By analyzing keyword frequencies as recorded in a citations database, one can characterize their replication in terms of a continuum between (a) random copying of fashionable buzzwords at one extreme (akin to random genetic drift), and (b) independent selection of keywords, based on inherent qualities, at the other (falsifying the neutral model). The question is one of degree, with variation expected along this basic continuum. Using random copying as the null hypothesis, one can simply seek to identify selection against the null without characterizing it specifically; although clearly the first hypothesis is that words are selected for usefully describing something real and relevant to the topic.
\\ \indent It may seem cynical to assume first that keywords are copied without much thought, but several studies suggest this [2,3,9,12] and even George Orwell thought as much in his famous 1946 essay, `Politics and the English language.' As the null hypothesis, random copying does not mean that the words themselves are chosen randomly, but that they are copied randomly from others who have already used them. The assumption is that randomly-copied keywords are  \textit{value-neutral}, in that no keyword is inherently more valuable than any other - the likelihood of any being chosen is simply proportional to its current popularity.  This is in essence the neutral model of population genetics  \cite{Gillespie_2004, Nowak_2006}.
\\ \indent In previous simulations, the random copying, or neutral, model has been represented as follows: Start with a set of $N$ individuals, which are replaced by $N$ new individuals in each generation.  Over successive generations, each of the $N$ new individuals copies its variant from a randomly-selected individual in the previous generation, with exception of a small fraction, $\mu$ $(< 5 \%)$, of the $N$ new individuals who invent a new variant in the current generation. 
\\ \indent The neutral model is simple to simulate, yet has been shown to provide richly complex results that produce at least three useful predictions relevant to cultural drift \cite{Hahn_Bentley_2003, Bentley_etal_2004, Bentley_etal_2007}: 

\begin{enumerate}
\item 
If individual variants are tracked through the generations, their frequencies (relative popularities) will change in a stochastic manner, as opposed to a directed manner or completely random manner.  More specifically, the haploid neutral model predicts that the only source of change in variant frequencies over time is random sampling, such that (3):\begin{equation}
V=\frac{\nu(1-\nu)}{N},
\end{equation}
where $V$ is the variance in frequencies from one time step to the next, and $\nu\leq1$ is the relative frequency of the variant as fraction of $N$, the maximum possible number of variant copies per generation. For small $\nu$, $\nu(1-\nu)\simeq\nu$, which after rearranging eq. (1) indicates that $NV/\nu\simeq 1$.
\item Like many processes of proportional advantage (under random copying the chance of being copied is proportional to current frequency), the variant frequencies exhibit a long-tailed distribution, which for small values of $\mu$ follows a power law form \cite{Bentley_etal_2004, Hahn_Bentley_2003}. This is one of the less diagnostic predictions, as a variety of mechanisms can generate power law and related distributions  \cite{Newman_2005}. Nonetheless, the distribution is useful as a null expectation. Among the possible departures from this null, selective bias for novelty (e.g., some maximum threshold of popularity) should truncate the tail (high end) of the variant frequency distribution \cite{Bentley_Shennan_2003, Mesoudi_Lycett_2008}.  Alternatively, there might be a conformist bias resulting in a `winner take all' distribution, whereby one word has a higher frequency than predicted by the power law for the rest of the words.
\item There is continual turnover in the variant pool. If the variants are ranked in order of decreasing frequency, the turnover $z$ in that list over successive generations (time) depends much more strongly on $\mu$ than on $N$ \cite{Bentley_etal_2007}, such that: \begin{equation}
z\approx\sqrt{\mu}
\end{equation}
where $z$ is measured as the fraction of turnover in the list (e.g., two items replaced in a Top 10  list would be 20\% turnover). In contrast to random copying, under selection the population size $N$ should correlate positively with the turnover rate in the ranked list of most popular variants  \cite{Bentley_etal_2007}.
\end{enumerate}	
	Using these three predictions as the null model, it is easier identify selection, which is effectively demonstrated by departures from these patterns, dependent on the kind of selection operating.
\\ \indent In applying this to keyword use, let $N$ represent the number of keywords in a given time period (rather than the number of articles, which vary in their number of keywords).  This ensures that each ÔindividualÕ corresponds with exactly one variant. The invention rate $\mu$ is then the fraction of those words in each time interval that are appearing for the first time.  

\section{Data}
The data used in this analysis were taken from Thompson Scientific's `Web of Science' (WoS) database, which covers articles thousands of journals in science and engineering, social sciences, arts and humanities. Among the wealth of information provided, each journal article description in the WoS database contains the title, keywords and abstract, references cited, and a list of all papers in other journals that have cited the paper to date.
\\ \indent As listed on the WoS database, the four case studies presented here provide a test of differences of keyword use among published articles within older paradigms versus younger ones, and within the physical sciences versus the social sciences. In order to define these case studies, we need a working definition of a sub-field of academic publishing. If belabored, this could be quite a difficult task -- many definitions would be too subjective, variable or broad. 
\\ \indent A way forward is to define a scientific `paradigm'  \cite{Kuhn_1962} as comprising the scientific papers that were in some way inspired by a certain highly-influential paper.  We thus can define each academic paradigm as the set of all papers that cited a certain highly-cited paper.  The citing papers may occur in a range of different journals, but they will all share the defining characteristic of citing the highly-influential work.
\\ \indent Consider four highly-cited, seminal works, two from the natural sciences and two from the social sciences.  To see the effect of time, from the pair in each category we include one work about 30 years old and the other about ten years old. This provides two comparisons: older versus younger fields of study, and social sciences versus physical sciences.  
\\ \indent From the physical sciences we have a paper by Barab\'{a}si and Albert (PS99, for `physical sciences, 1999') in 1999 \cite{Barabasi_Albert_1999}, which introduced a quantitative model of `scale-free networks' and has been cited over 2,000 times (as listed on the WoS database), and one by Witten and Sander (PS81) from 1981 \cite{Witten_Sander_1981}, which introduced the physics model of  `diffusion limited aggregation', and has been cited over 1,300 times. From the social sciences, there is a paper by Nahapiet and Ghoshal (SS98) in 1998 \cite{Nahapiet_Ghoshal_1998}, cited over 460 times, which reviewed the influential concept of `social capital', and a 1977 book by Bordieu (SS77), cited over 2,700 times, which introduced such influential concepts as `agency' and `structuration' into the social sciences \cite{Bordieu_1977}.
\\ \indent For each of the sets of articles within each defined paradigm, the keywords data from the WoS database were taken only from titles and keywords chosen by the authors (not the WoS `Keywords plus' which is an automated condensation of the cited references), and then sorted by publication year. The following common words were removed from the data: \textit{a, an, and, as, by, for, from, in, its, of, on, the, to, using}, and \textit{with}. Aside from these, no other common words were present in high enough frequencies to significantly affect the patterns discussed below. 
\section{Results}
Figure 1 shows the temporal change in $N$, the number of keywords for each case study per year, and in $N\mu$, the number of new keywords per year, for paradigms about 10 years old (Figure 1a) and 30 years old (Figure 1b). A new keyword was one which had not appeared in the record beforehand, with records starting in 1994 for the older works and date of publication (1998, 1999) for the younger paradigms.  
\\ \indent Table 1 shows additional statistics for each paradigm averaged from 2002 to 2006, the sample period applicable to all four case studies (the newer case studies do not have enough data before 2002). In each case, the quantities $N$ and $N\mu$ parallel each other (Figure 1), indicating a consistent and substantial invention rate  $\mu$ between 15 and 30\% in all cases (Table 1). Within the older pair and the younger pair of paradigms, the invention rate $\mu$ was higher for the social science than for the physical science case (Table 1). This is true even though the comparison differs in the number of words: $N$ is larger for PS99 than SS98, but lower for PS81 than SS77.

\begin{table}[htbp]
\caption{\textbf{Average values, from 2002-2006,} of the number of keywords $N$, newly appearing keywords $N\mu$, and different keywords or `vocabulary'.  The invention fraction  $\mu$ is shown as a range, representing the decline in this value over the time period.}
\begin{center}
\begin{tabular}{lcccc}
\ &SS77&PS81&PS99&SS98\\\hline $N$&1671&1050&2660&885\\Vocab&1036&566&979&431\\$N\mu$&441&192&511&224\\$\mu$, \%&45-18&28-16&48-13&52-14
\end{tabular}
\end{center}
\end{table}
\begin{figure}
\begin{center}
\includegraphics[width=1.35in]{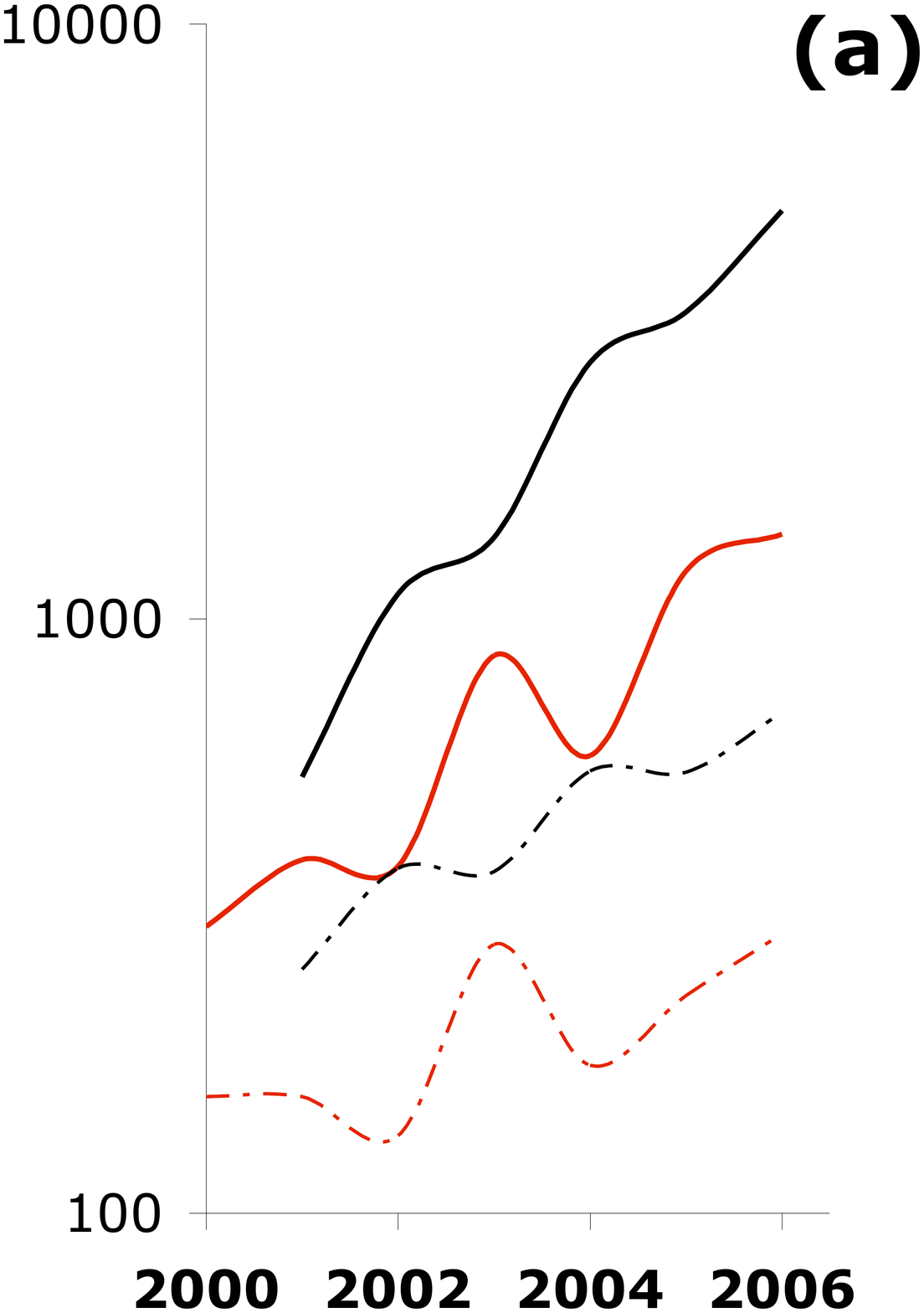}
\includegraphics[width=1.35in]{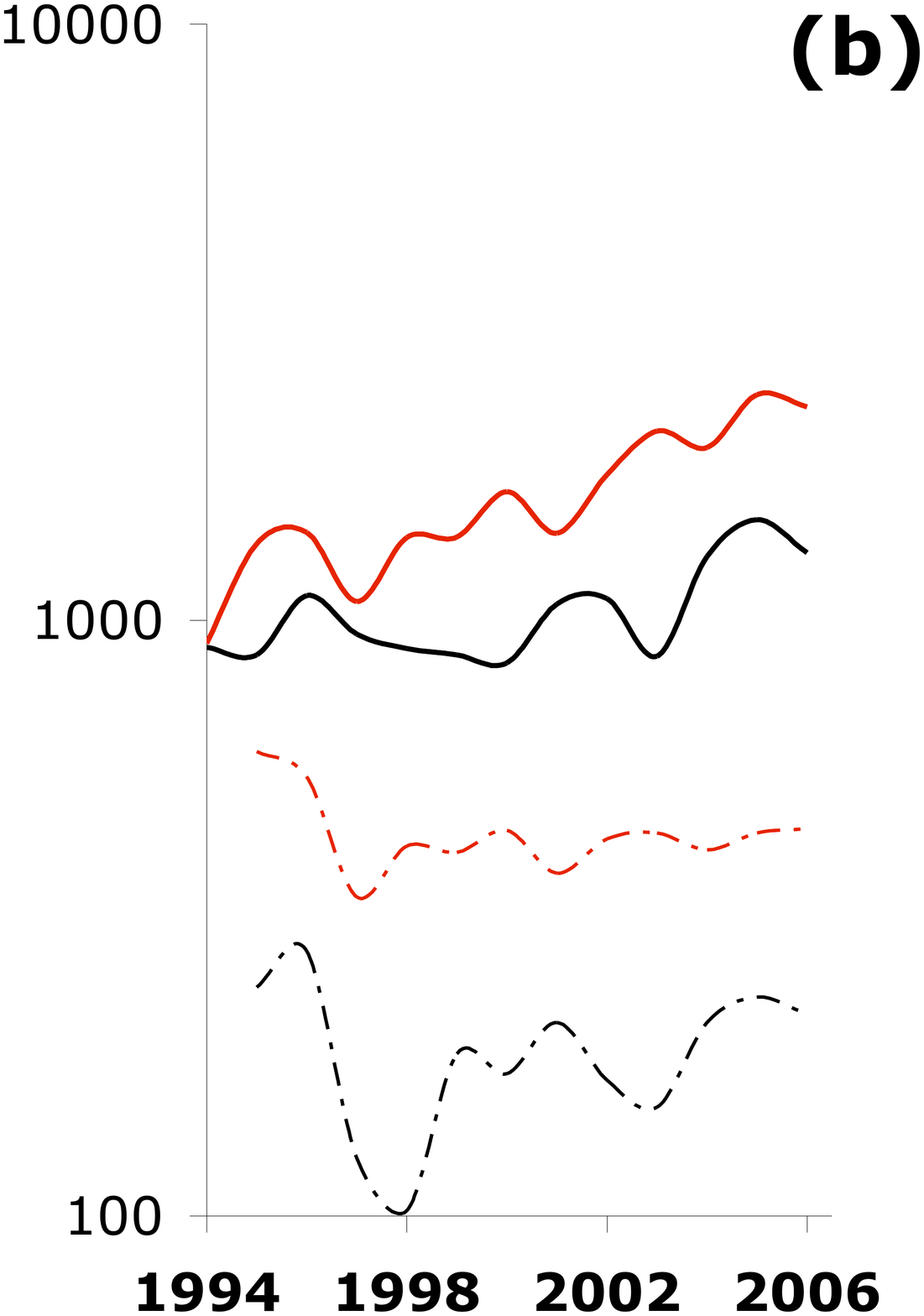}
\end{center}
\caption{\textbf{Keywords, total and new, among paradigms about (a) 10 and (b) 30 years old.} Social science cases \textcolor{red}{in red} and physical sciences \textbf{in black}. Solid curves show the total number of keywords $N$ per year, and the dashed curve shows number of new keywords $N\mu$ introduced per year. Logarithmic y-axis.}
\end{figure}

\paragraph{} In addition to a higher innovation fraction for the social science paradigms, there is also a marked difference in the turnover in keywords. Consider the top 5 keywords, in terms of popularity, over the years in each case study (below the top 5, keywords start to become insufficient in their numbers of appearances). As the best way to view overall trends in turnover, Figure 2 shows the cumulative turnover in the top 5 keywords, expressed as a fraction (e.g., 4 words having passed through the top 5 = 80\% turnover). In the physical science paradigms, the turnover in the top 5 keywords leveled off to virtually no turnover in the last several years. At the other end of the spectrum, the keywords in the social science cases show a high and steady turnover throughout the sampling period (Figure 2). In the case of SS77, this turnover persisted long after its publication, and many years beyond which PS81 had leveled off.
\begin{figure}
\begin{center}
\includegraphics[width=2.7in]{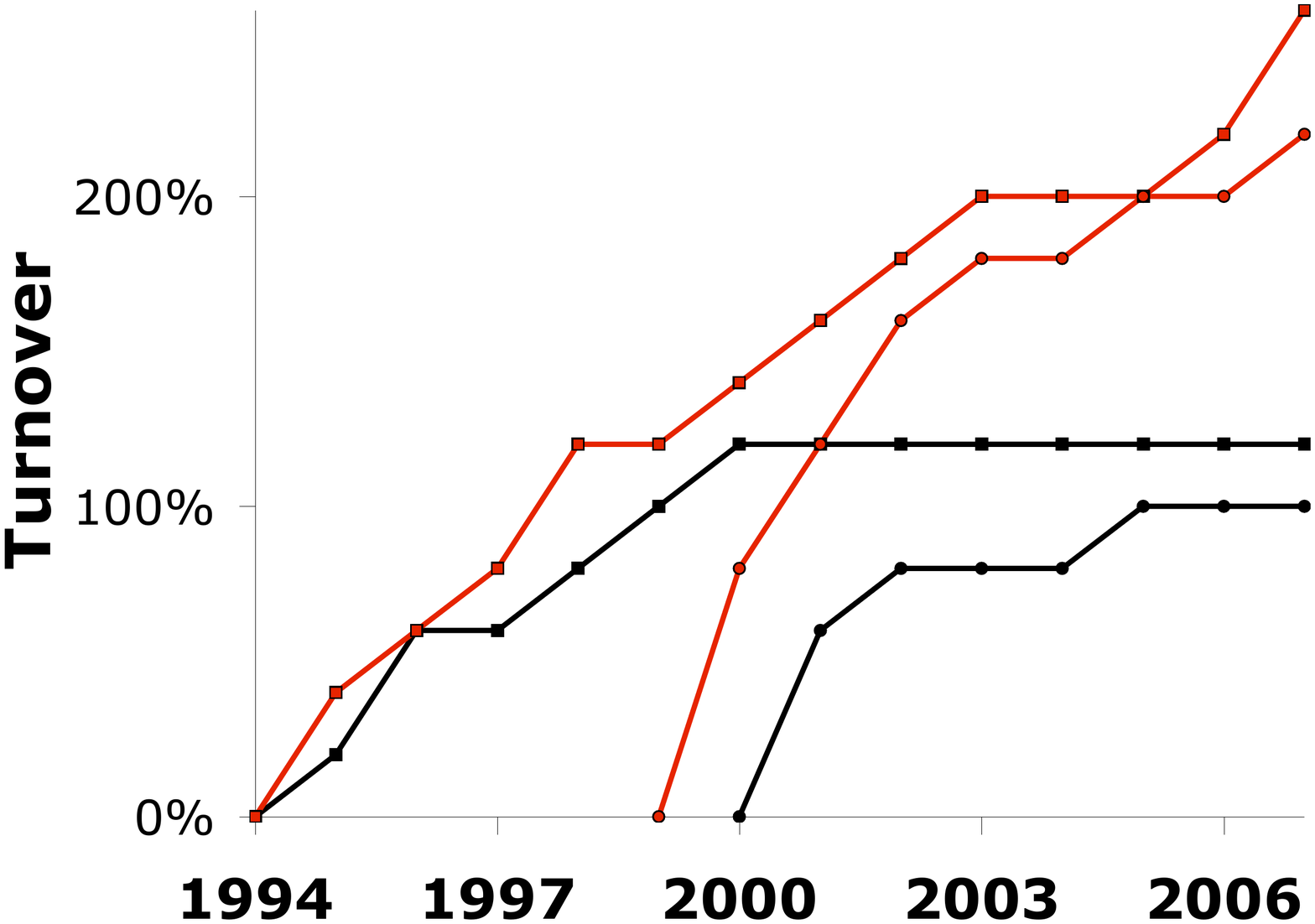}
\end{center}
\caption{\textbf{Cumulative turnover in the top 5 keywords.} Social science cases shown \textcolor{red}{in red} and physical sciences \textbf{in black}. Turnover refers to words making a first appearance in the top 5. For the older paradigms (SS77; PS81), symbols are squares and the count begins at 1994, for the newer articles (PS99; SS98) symbols are circles and the count begins the year after publication.}
\end{figure}
\\ \indent Whereas the continual turnover in SS77 and SS98 is consistent with random copying with innovation, the cessation of turnover in PS81 and especially PS99 suggests selection. As Figure 3a shows, the selective sorting of the keyword frequencies for PS99 was strong enough that even the keyword \textit{networks} (highlighted in red) occupies a distinct frequency ranking from the singular \textit{network} (blue), while other entries are similarly locked into their positions among the top 5. Although this pattern of selection is not as strong in the older physical science paradigm (PS81), the blue versus black lines in Figure 3c show apparent groupings of words by selected frequencies. In contrast, both the older and younger social science cases (SS77 and SS98) appear more stochastic in their histories of individual word frequencies (Figure 3 b, d), and with each at a relatively low frequency compared to the network science case (Figure 3 a, c). In the SS77 case, the ratio $NV/\nu$ increases moving down the rankings (Table 2), which suggests a possible conformist bias, in that the more frequent words have been preferentially selected (e.g. red curve in Figure 3d).
\\ \indent As described above, the ratio $NV/\nu$ can be used to characterize keyword variability, allowing comparison across cases studies for the period 2002-2006 (Table 2). Averaged over the five keywords, $NV/\nu$ differs more by age of the paradigm than by subject matter, being higher for the younger (2.3) than the older (1.3-1.4) paradigms. Within each age pair, however, the physical sciences paradigm has the larger standard error in the mean value of $NV/\nu$ (Table 2). This reflects certain keywords in the physical science paradigms whose popularity changed directionally, apparently due to selection. In the PS99 case, the word \textit{complex} (word 3, $NV/\nu$ = 5.8) appears to have been selected for, as it doubled in frequency from 2002 to 2006 beyond what would be expected from random drift. Also in the PS99 case, \textit{networks} (word 1) declined steadily as \textit{network} (word 2) increased, such that their variability scores are near 2. By contrast, the words in the SS98 case do not show such directionality in their change (Figure 3b), and the high variability scores for four of the five words (Table 2) is due to their fluctuating frequencies over the time interval (Figure 4b). Curiously, in the PS81 case, the word \textit{aggregation} (word 2, $NV/\nu$ = 0.3) was considerably less variable than \textit{diffusion} (word 5, $NV/\nu$ = 2.7) even though the seminal paper \cite{Witten_Sander_1981} was about diffusion-limited aggregation.  
\begin{table}[htdp]
\caption{\textbf{Values of $NV/\nu$ for the top 5 words, 2002-2006}, tracked in Figure 3. Numbers in parentheses give standard error on the trailing digits.}
\begin{center}
\begin{tabular}{lcccc}
\   &SS77&PS81&PS99&SS98 \\\hline Wd1&0.82&1.52&2.10&2.73\\ Wd2&1.13&0.34&1.82&2.75\\ Wd3&1.10&0.71&5.85&0.97\\ Wd4&1.46&1.55&1.28&2.66\\ Wd5&1.75&2.69&0.59&2.19\\ \bf Ave&\bf1.25(16)&\bf1.36(41)&\bf2.33(92)&\bf2.26(34)
\end{tabular}
\end{center}
\end{table}

\begin{figure}
\begin{center}
\includegraphics[width=2.2in]{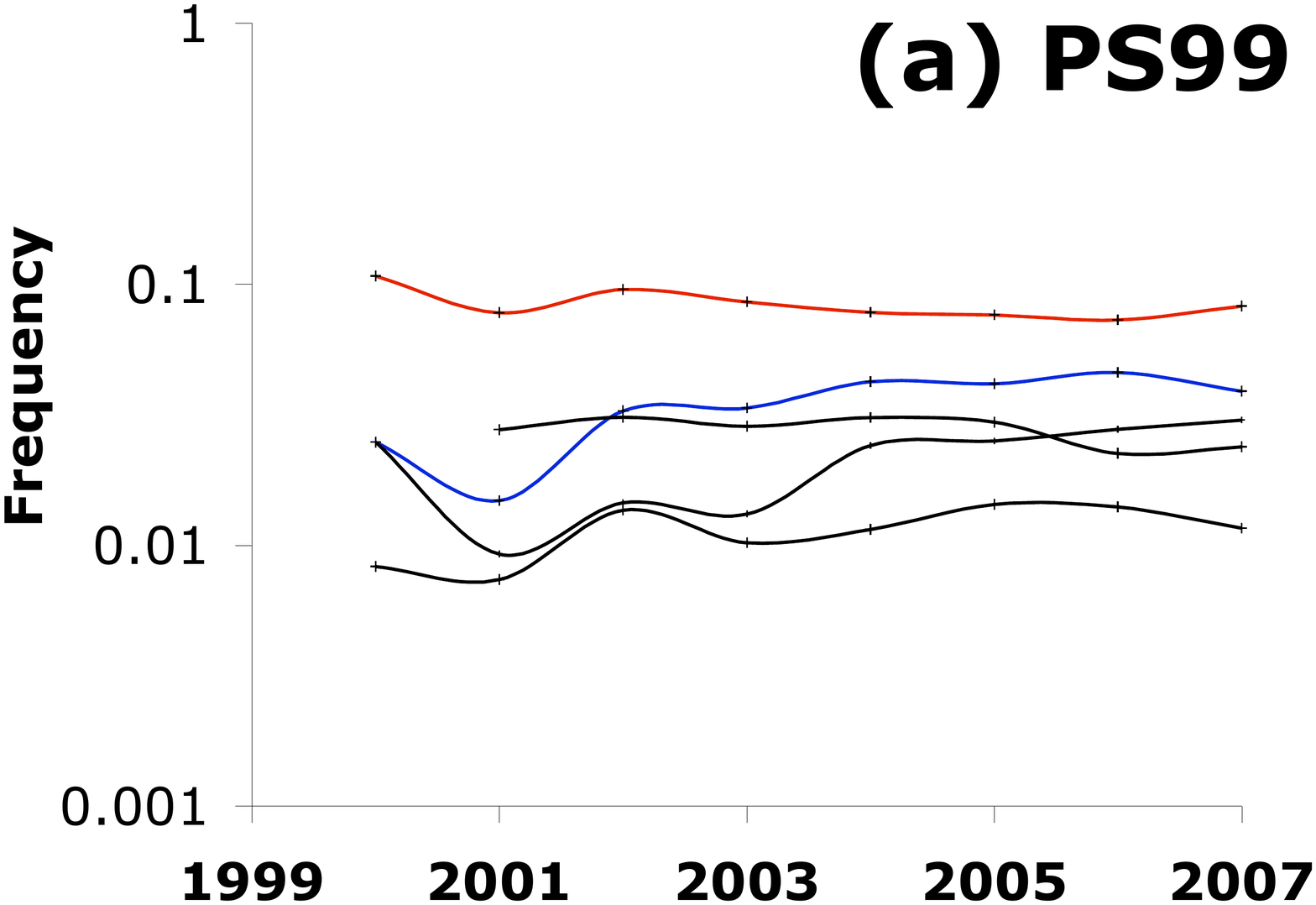}
\includegraphics[width=2.2in]{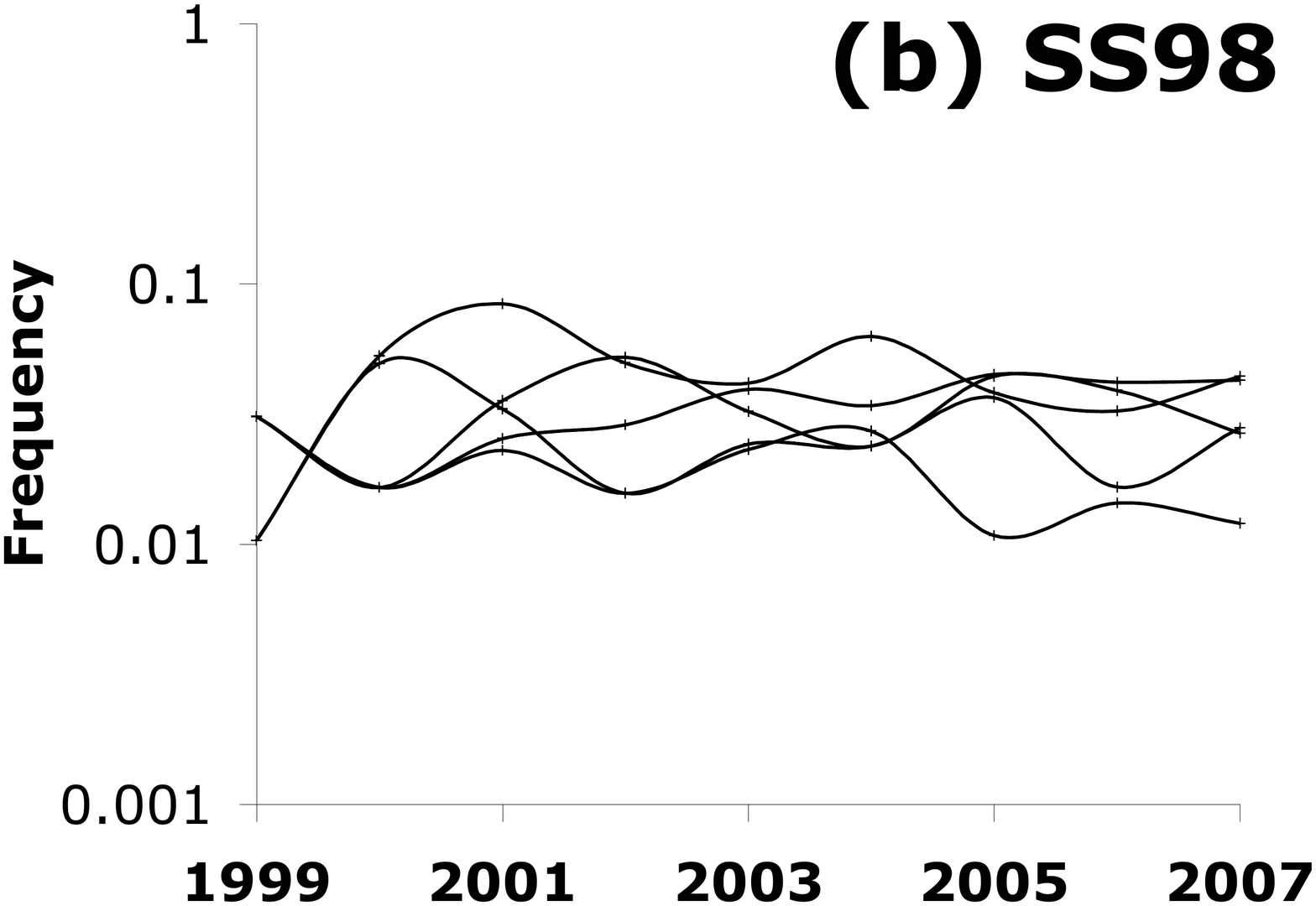}
\includegraphics[width=2.2in]{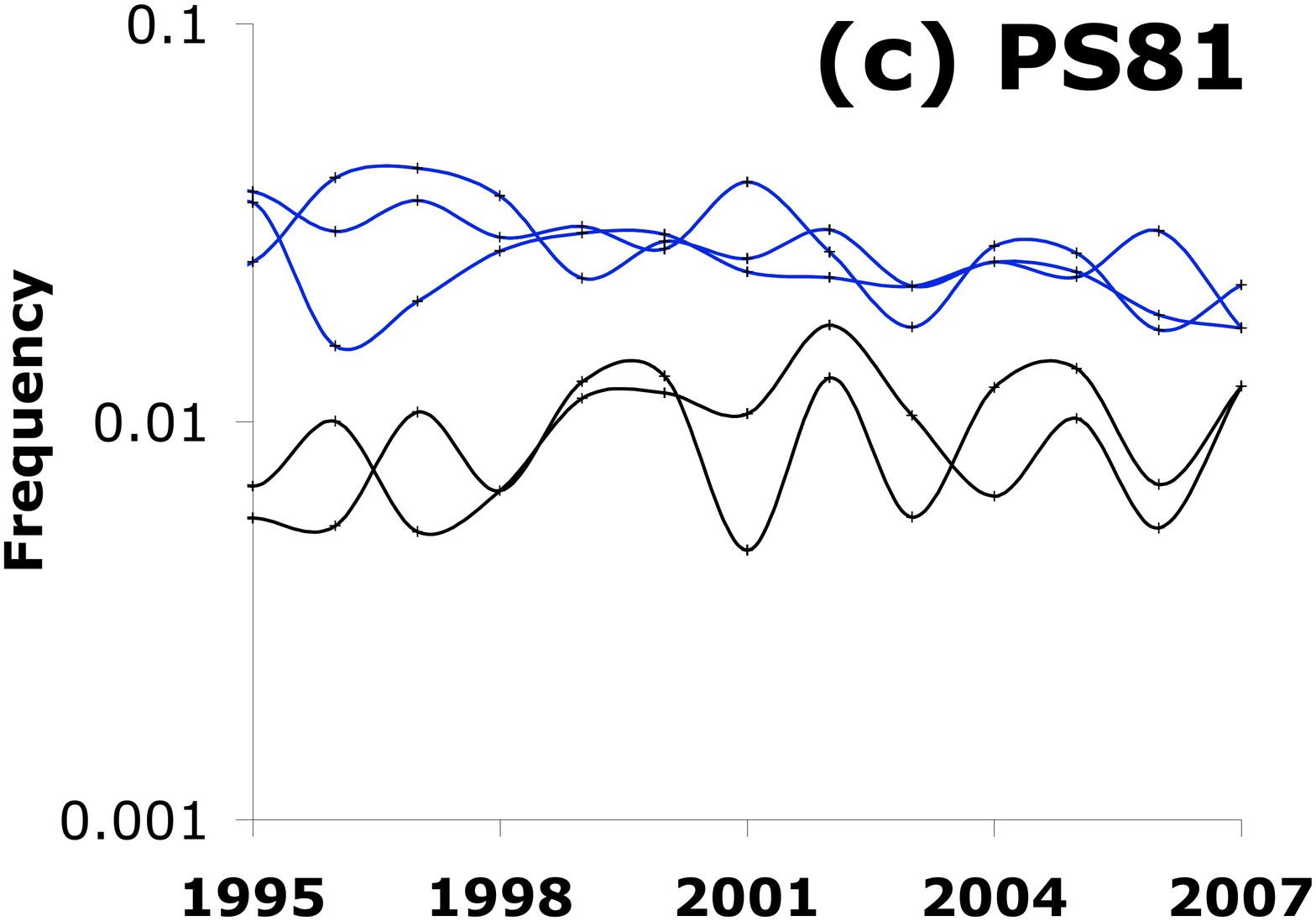}
\includegraphics[width=2.2in]{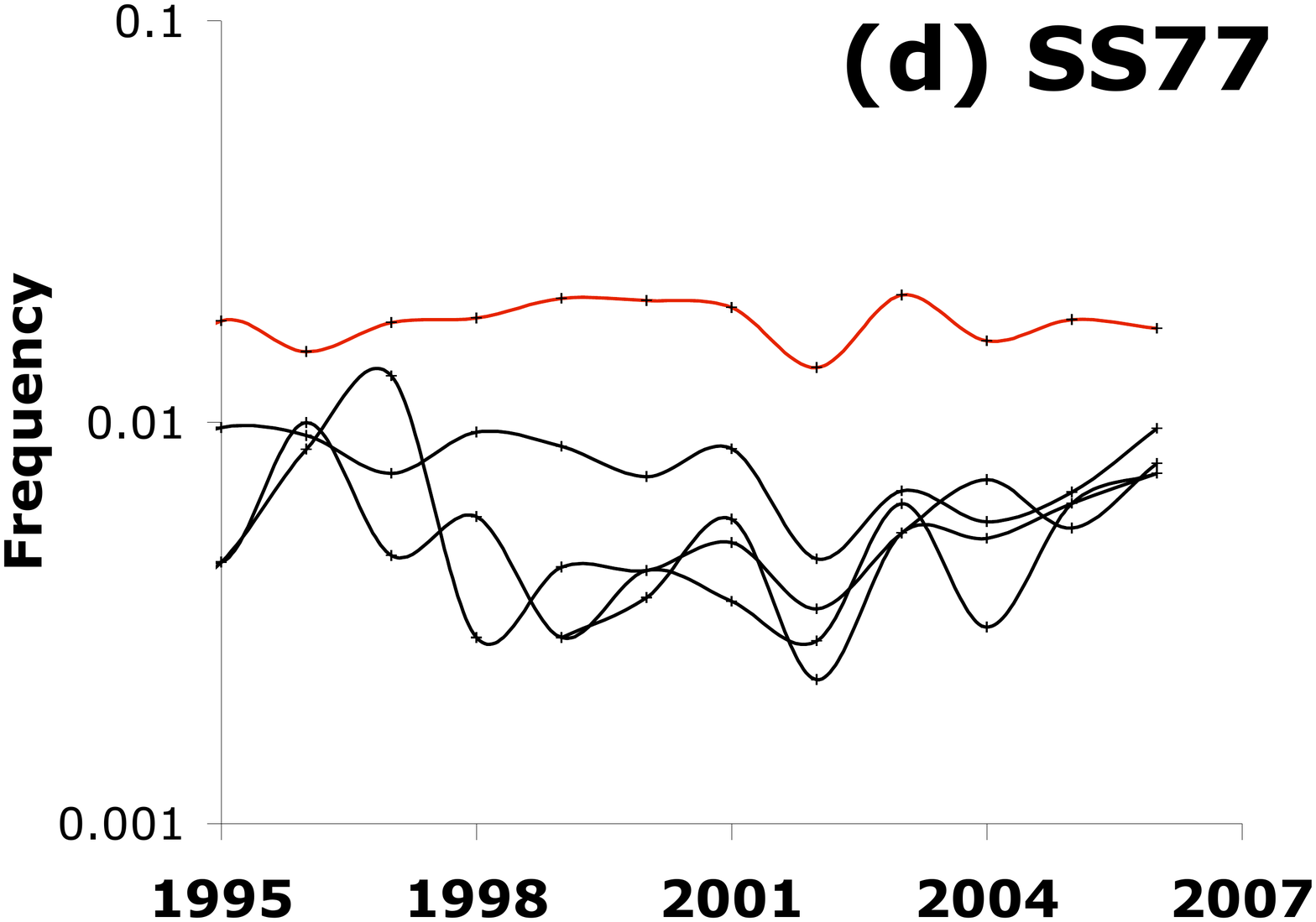}
\end{center}
\caption{\textbf{Frequencies of the top 5 keywords of 2005.} Shown are the four paradigm case studies, including: (a) newer physical sciences (PS99); (b) newer social sciences (SS98); (c) older physical sciences (PS81); and (d) older social sciences (SS77). Logarithmic y-axes.}
\end{figure}
\begin{figure}[htbp]
\begin{center}
\includegraphics[width=2.1in]{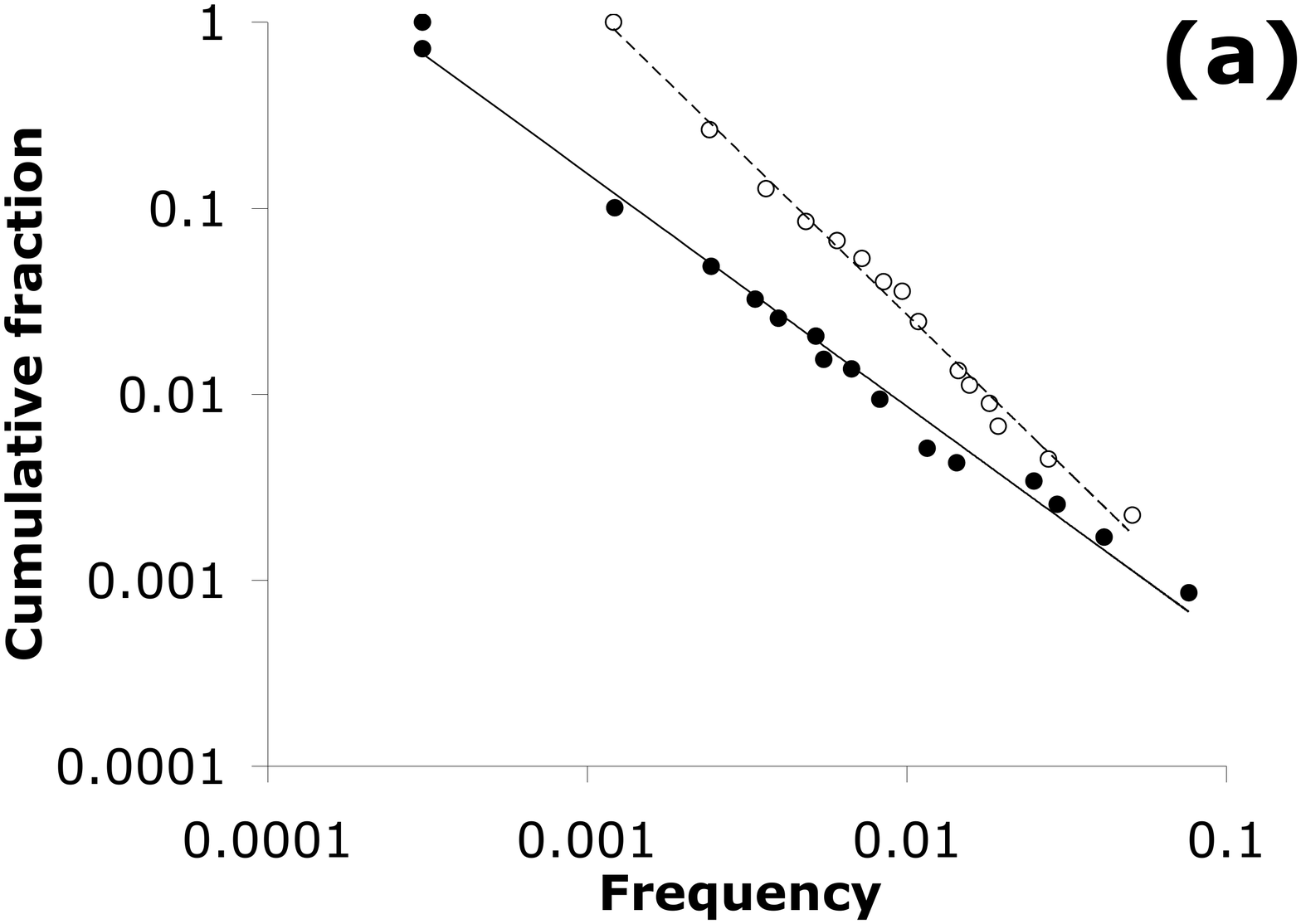}
\includegraphics[width=2.1in]{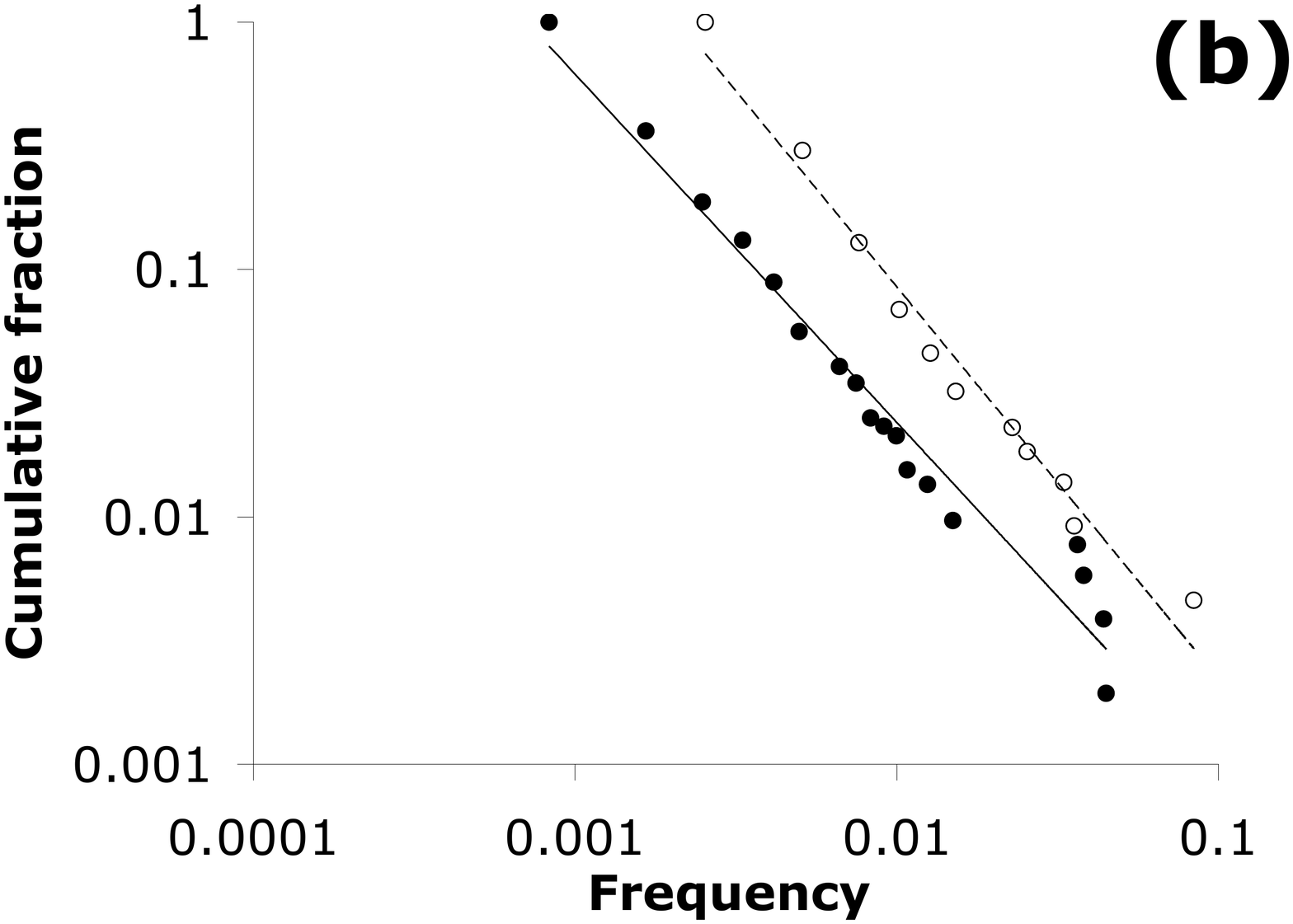}
\includegraphics[width=2.1in]{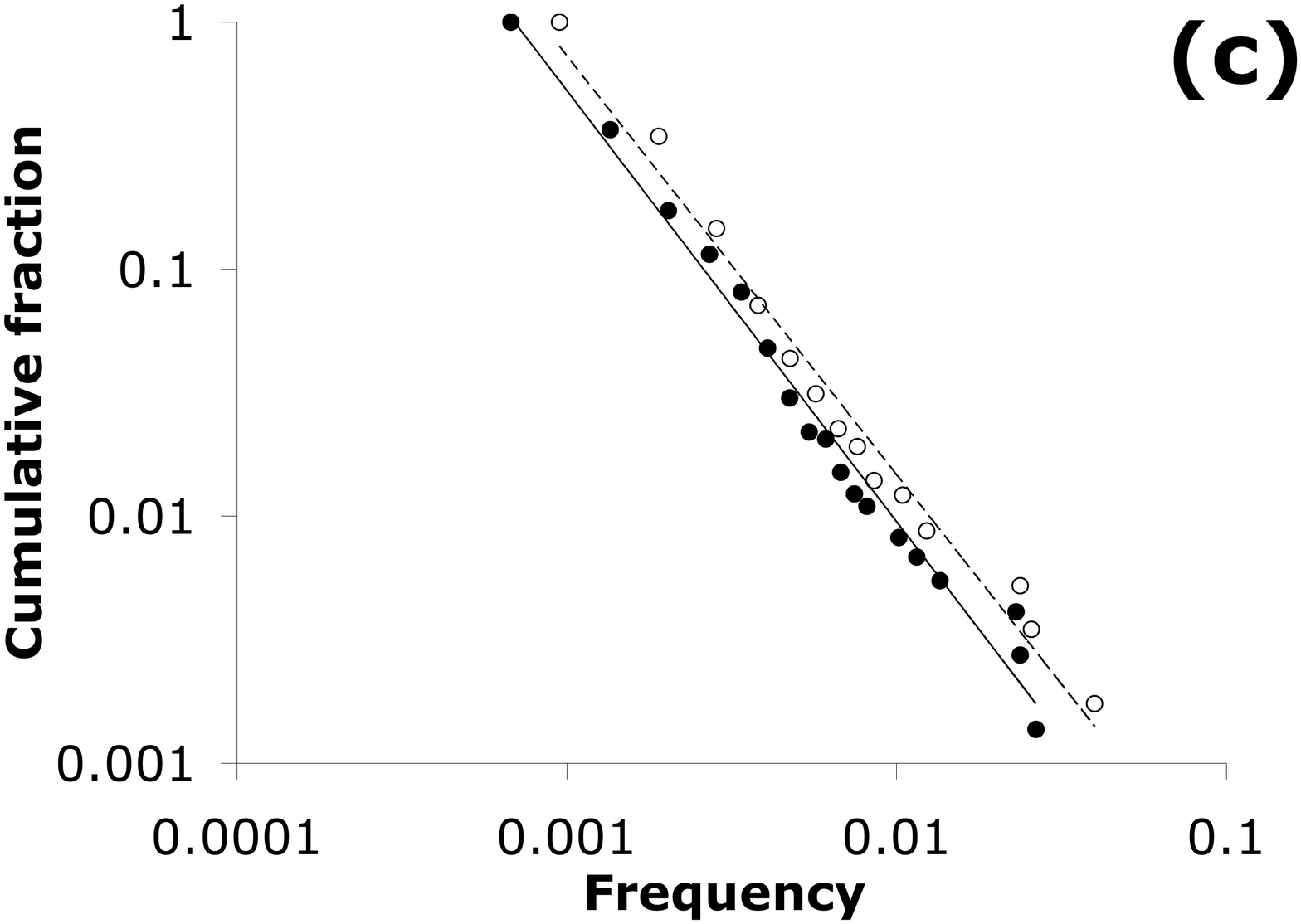}
\includegraphics[width=2.1in]{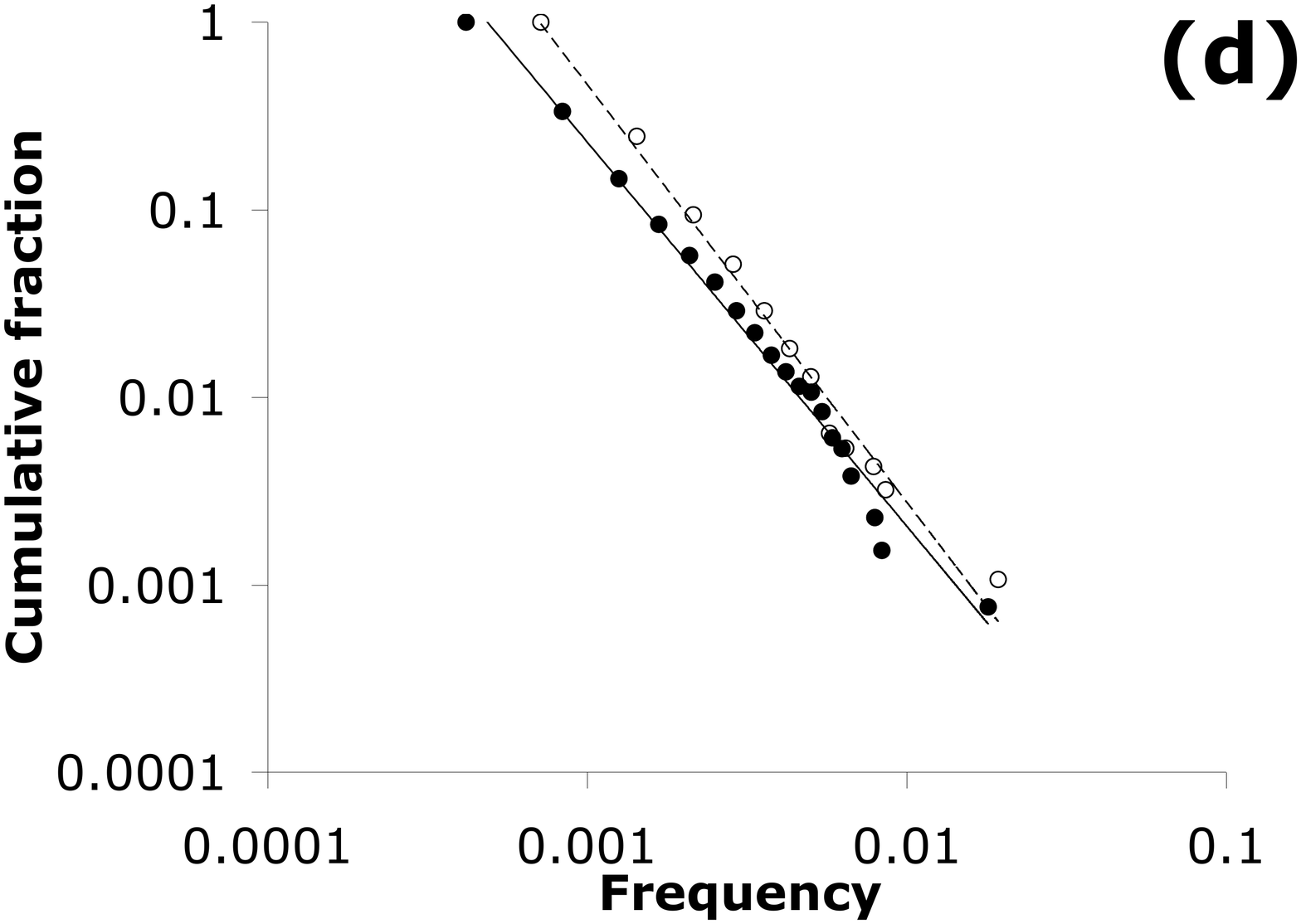}
\end{center}
\caption{\textbf{Cumulative frequency distributions of all keywords.} Open circles show distribution for 2001 and filled circles are for 2005. The paradigms are (a) newer physical sciences (PS99), (b) newer social sciences (SS98), (c) older physical sciences (PS81) and (d) older social sciences (SS77). Using the least-squares method \protect\cite{Newman_2005}, the estimated power-law exponents for 2001 and 2005, respectively, are as follows: PS99: 2.11, 2.00; SS98: 2.11, 2.02; PS81: 2.09, 2.05; SS77: 2.18, 2.09. Errors (by jackknife estimate) on these exponents are $< 0.01$.}
\end{figure}

\paragraph{} Finally, consider keyword frequency distributions for two time-slices, years 2001 and 2005 (Figure 4).  All show essentially a power law form, which could be consistent with either the neutral model but also a variety of models of proportionate advantage \cite{Newman_2005}.  More revealing is the degree of change in the power law exponent (slope on the log-log plot) over this 4-year time span. In three cases, the slope is nearly the same for 2005 as for 2001, but for PS99, the slope is considerably less for 2005. The decreasing slope for PS99 correlates with a decreasing invention rate $\mu$ over this time span (Table 1), and reflects the diminishing probability for any new keyword to ever reach the top 5.  
\\ \indent The frequency distributions in Figure 4 enable the identification of copying biases. Although all four paradigms yield essentially power law distributions, in two cases -- PS81 and SS98 -- show marked departures from a power law in the truncations of the tail (Figure 4b and 4c).  In each case there appears to be selection for the top 3 or 4 words (and they are the same words in 2001 and 2005 for each case), such that their frequencies are roughly the same rather than following the power law.

\section{Discussion \& conclusions}
By treating academic keywords as discrete elements of evolution, this study finds that different academic niches -- as defined by sets of publications which share a single seminal article in their cited reference lists -- can show markedly different evolutionary patterns. From the case studies considered, it appears that some academic fields are characterized by a high degree of drift, resulting in continual and unpredictable change in vocabulary, whereas in others words appear under selection, such that the predominant vocabulary becomes increasingly crystallized and unchanging over time.  
\\ \indent Among the cases presented, the social science paradigms showed the stronger patterns of random copying with invention, including constant turnover in the keywords of highest frequency, and the stochastic ups-and-downs of individual word frequencies over time. In contrast, the physical science paradigms showed a rejection of the neutral model, particularly in the cessation of turnover in the top keywords over time.  
\\ \indent The scale of analysis is a key variable; a certain group of keywords might be selected, yet drifting within the group. Similarly, in a different study, while choices of baby names for the entire United States are indistinguishable from random copying \cite{Hahn_Bentley_2003}, different ethnic groups certainly select from different pools of names \cite{Fryer_Levitt_2004}, and it remains to be studied whether random drift would predominate again within these groups. 
\\ \indent In addition to these particular points, this study is meant to demonstrate how a similar evolutionary analysis could be performed on any cultural dataset comprising discrete elements. This evolutionary analysis contrasts with the increasing representation of knowledge growth as networks \cite{Guimera_etal_2005, Wuchty_etal_2007, Humphries_etal_2008} with the individuals (e.g. authors) as `nodes', and their interactions (e.g. cited references) as `links'. A particular challenge for network analysis, however, is change, because a network implies a structure to interactions -- the connections of today determine what will happen tomorrow, such that change must be modeled as a modification of the existing network. However, in fashionable realms, yesterday might be less important than tomorrow, and interactions of influence may differ completely from one day to the next. Change can be the essence of the process, rather than just a modification.
\\ \indent For this reason, evolutionary theory can often naturally account for change that may be seen as exceptional in a network model \cite{Minnhagen_Bernhardsson_2008, Humphries_etal_2008, Volz_Meyers_2008}. A recent network analysis \cite{Palla_etal_2007}, for example, tracked coauthorships and mobile phone calls to show that, in order to have longevity, small groups require stability in their composition, whereas large groups last a bit longer with some degree of turnover in their membership. This is, in fact, a basic prediction of the genetic drift model: small populations are destroyed by drift, large populations can tolerate it and even find it adaptive. The crucial difference is that in the network analysis \cite{Palla_etal_2007} mutation was measured as absolute number of changes, whereas the random copying model defines mutation $\mu$ as a fraction of $N$. Hence the random copying model would have predicted the network result, in that coherence disintegrates more quickly with one mutation per time step in a population of 4 versus a population of 100, for example, because the former is a much higher mutation rate.
 \\ \indent Change, in fact, is central to evolutionary theory. The use of some basic evolutionary analyses, with parallels in population genetics, can be used to characterize different forms of innovation and transmission of discrete cultural elements. Identifying what proceeds in predictable directions, as opposed to drifting upon the tides of fashion, would be of great utility in understanding the evolution of knowledge. It is wasted effort to try to predict the future of randomly drifting fashionable buzzwords \cite{Bentley_2006, Salganik_etal_2006}, but one might hope to predict selected elements, such as valid new scientific terms. The kind of evolutionary analysis used here is generally applicable to any case study where popularity can be presented in the form of frequencies and ranked lists over time.
\bibliographystyle{abbrv}

\end{document}